# Heterogeneous anomalous diffusion in view of superstatistics

Yuichi Itto

*Science Division, Center for General Education, Aichi Institute of Technology,*

*Aichi 470-0392, Japan*

**Abstract**   It is experimentally known that virus exhibits stochastic motion in cytoplasm of a living cell in the free form as well as the form being contained in the endosome and the exponent of anomalous diffusion of the virus fluctuates depending on localized areas of the cytoplasm. Here, a theory is developed for establishing a generalized fractional kinetics for the infection pathway of the virus in the cytoplasm in view of superstatistics, which offers a general framework for describing nonequilibrium complex systems with two largely separated time scales. In the present theory, the existence of a large time-scale separation in the infection pathway is explicitly taken into account. A comment is also made on scaling nature of the motion of the virus that is suggested by the theory.

1. Introduction

In recent years, the infection pathway of adeno-associated viruses in living *HeLa* cells has experimentally been studied by making use of the technique of real-time single-molecule imaging [1,2]. (Here, the adeno-associated virus is a small virus particle, and the *HeLa* cell is a line of human epithelial cells.) Remarkably, an exotic phenomenon has been observed in cytoplasm of the living cell. In the experiments [1,2], the virus solution of low concentrations was added to a culture medium of the living cells. Then, the trajectories of the viruses, each of which is labeled with fluorescent dye molecule, in the cytoplasm were analyzed. The experiments show that the fluorescent viruses exhibit stochastic motion in the free form as well as the form being contained in the endosome (i.e., a spherical vesicle).

Let us denote the mean square displacement in stochastic motion by $\overline{x^2}$. It behaves as

$$\overline{x^2} \sim t^\alpha, \qquad (1)$$

for large elapsed time, $t$, in general. Normal diffusion leads to $\alpha = 1$, while $0 < \alpha < 1$ ($\alpha > 1$) corresponds to subdiffusion (superdiffusion). The experimental result shows that the mean square displacement of the fluorescent virus exhibits not only normal diffusion but also subdiffusion in the form of Eq. (1). However, what is truly remarkable is the fluctuations of $\alpha$ in the case of subdiffusion [1]: $\alpha$ fluctuates between 0.5 and 0.9, depending on localized areas of the cytoplasm. This may be due to existence of obstacles in the cytoplasm, not the forms of existence of the virus (i.e., being free or contained in the endosome) [1,2]. Thus, this phenomenon is seen to manifest the *heterogeneous* structure of the cytoplasm as a medium for stochastic motion of the virus.

The phenomenon mentioned above is in marked contrast to traditional anomalous diffusion [3] discussed for a variety of physical systems, examples of which are particle motion in turbulent flow [4], charge carrier transport in amorphous solids [5], the flow of contaminated vortex in fluid [6], chaotic dynamics [7], porous glasses [8]. On the other hand, in biology, a lot of efforts have been devoted to understanding the virus infection process in order to both design antiviral drug and develop efficient gene therapy vectors.

The purpose of this paper is to develop a theoretical framework for establishing a generalized fractional kinetics proposed in Ref. [9], where the infection pathway of the adeno-associated virus in the cytoplasm of the living *HeLa* cell is studied by generalizing traditional fractional kinetics [10]. For this purpose, we base our consideration on the idea of superstatistics. Superstatistics, which has been introduced in Ref. [11] after some preliminary works in Refs. [12-14], is "statistics of statistics" with two largely separated time scales and offers a unified theoretical framework for describing nonequilibrium complex systems with two such time scales. A prototype system in superstatistics [11,13] is a Brownian particle moving through a fluid environment with varying temperature on a large spatial scale. This system is divided into many small spatial "cells", each of which is in local equilibrium characterized by each value of temperature. So, the Brownian particle in a given cell moves to neighboring ones. Then, variation of the local fluctuations of temperature is slow, whereas relaxation of the Brownian particle in a cell to a local equilibrium state with a given value of temperature is fast. Consequently, the system on a long time scale is described by a superposition of two statistics associated with these different dynamics. The situation we consider here is that the virus moves through the cytoplasm with varying local fluctuations of the exponent, $\alpha$, in Eq. (1). We assume that there is a large separation of two time scales in the infection pathway: the time scale of variation of exponent fluctuations is much larger than that of stochastic motion of the virus in a localized area of the cytoplasm. This is in analogy with the existence of two largely separated time scales in superstatistics. From the viewpoint of superstatistics, we describe the motion of the virus by a superposition of two different statistics: one is statistics concerning stochastic motion of the virus in a localized area of the cytoplasm, and the other is one associated with variation of exponent fluctuations. For the virus in each localized area, we apply fractional kinetic theory, which generalizes Einstein's approach to Brownian motion [15]. Proposing the statistical form of the fluctuations of the exponent based on the experimental data as well as the maximum entropy principle [16], we show that the present framework yields the generalized fractional kinetic theory. We also make a comment on a scaling law for the motion of the virus that is suggested by the theory.

**2. Stochastic motion of the virus in the cytoplasm and superstatistics**

Let us start our discussion with considering 1-dimensional stochastic motion of the virus in the cytoplasm with slowly varying local fluctuations of the exponent, $\alpha$. We describe it in view of superstatistics. To do so, we regard the cytoplasm as a medium for stochastic motion of the free virus as well as the virus contained in the endosome. Then, we imaginarily divide the medium into many small blocks, each of which is identified with a localized area of the cytoplasm. As already mentioned in the Introduction, the time scale of variation of the fluctuations is supposed to be much larger than that of stochastic motion of the virus in each local block. In other words, $\alpha$ is approximately constant while the virus moves through the blocks. For the virus in a local block with a given value of $\alpha$, we describe the probability of finding the virus in the interval $[x, x+dx]$ at time $t$ by $f_\alpha(x,t)dx$. Denoting the statistical distribution of the fluctuations of $\alpha$ by $P(\alpha)$, we describe the probability of finding the virus on a long time scale by the average of $f_\alpha(x,t)dx$ with respect to $P(\alpha)$:

$$f(x,t)dx = dx \int d\alpha\, P(\alpha) f_\alpha(x,t). \qquad (2)$$

Equation (2) clearly shows that the statistical property of the virus in the cytoplasm is given by the superposition of $f_\alpha(x,t)dx$ with respect to $P(\alpha)$ in conformity with the viewpoint of superstatistics.

In what follows, we first formulate a generalized fractional kinetic theory, in which the statistical fluctuation of the exponent is incorporated, based on Eq. (2).

We express $f_\alpha(x,t)dx$ in Eq. (2) in terms of $f(x,t)dx$ based on the scheme of continuous-time random walks [17]:

$$f_\alpha(x,t)\,dx = dx \int_{-\infty}^{\infty} d\Delta \int_0^t d\tau\, f(x+\Delta, t-\tau)\phi_\tau(\Delta)\psi_\alpha(\tau) + \delta(x)R(t)\,dx. \qquad (3)$$

Here, the first term on the right-hand side stands for all of possible probabilities that the virus moves into the interval from outside or stays in the interval. The second term is a partial source guaranteeing the initial condition, $f(x,0) = \delta(x)$, and $R(t)$ describes a time-dependent partial source with the condition, $R(0) = 1$. Then, $\phi_\tau(\Delta)$ is the normalized probability density distribution for a displacement, $\Delta$, in a finite time step, $\tau$. This distribution is sharply peaked at $\Delta = 0$ and satisfies the condition, $\phi_\tau(\Delta) = \phi_\tau(-\Delta)$. $\psi_\alpha(\tau)$ is the normalized probability density distribution for $\tau$, which is treated as a random variable, and satisfies the condition, $\psi_\alpha(0) = 0$. As will be seen below, it is implied that this distribution decays as a power law characterized by $\alpha \in (0,1)$ for long time step [see the discussion after Eq. (6) below]. From the normalization condition on $f(x,t)$, it is found that $R(t)$ is connected to $\psi_\alpha(\tau)$ through the relation: $R(t) = 1 - \int_0^t d\tau \psi_\alpha(\tau)$ [from which $R(t)$ depends on $\alpha$]. In our later discussion, we shall show how the present theory yields traditional fractional kinetics [10], which turns out to reproduce both normal diffusion and subdiffusion with a fixed exponent observed in the experiments.

Now, it seems that the nature of subdiffusion observed in the experiments comes from $\psi_\alpha(\tau)$, not $\phi_\tau(\Delta)$. Therefore, we assume in what follows that $\phi_\tau(\Delta)$ is actually independent of time steps: $\phi_\tau(\Delta) = \phi(\Delta)$. To formulate the generalized fractional kinetic theory, we employ the Laplace transforms of Eqs. (2) and (3) with respect to time:

$$\tilde{f}(x,u) = \int d\alpha\, P(\alpha)\, \tilde{f}_\alpha(x,u), \tag{4}$$

$$\tilde{f}_\alpha(x,u) = \int_{-\infty}^{\infty} d\Delta\, \tilde{f}(x+\Delta, u)\phi(\Delta)\tilde{\psi}_\alpha(u) + \delta(x)\frac{1-\tilde{\psi}_\alpha(u)}{u}, \tag{5}$$

where $\tilde{f}(x,u)$, $\tilde{f}_\alpha(x,u)$, and $\tilde{\psi}_\alpha(u)$ are the Laplace transforms of $f(x,t)$, $f_\alpha(x,t)$, and $\psi_\alpha(\tau)$, respectively, provided that $\mathcal{L}(g)(u) = \tilde{g}(u) = \int_0^\infty dt'\, g(t')e^{-ut'}$.

In analogy with the discussions in Refs. [18,19], where the separation of the time scales in superstatistics is explicitly implemented by the use of conditional concepts, we notice here the following point: in Eq. (4), the averaging over the slow variable, i.e., $\alpha$, is taken after the integration over the fast variable, i.e., $\Delta$, is performed. This procedure is opposite to that discussed in Ref. [9], where the integration over the fast variable is performed after the elimination of the slow variable. Thus, the existence of a large time-scale separation in the infection pathway is explicitly taken into account in the present procedure.

In Eq. (5), we suppose that $\tilde{\psi}_\alpha(u)$ takes the following form:

$$\tilde{\psi}_\alpha(u) \sim 1 - (su)^\alpha \tag{6}$$

with a characteristic constant, $s$, which has the dimension of time. This characteristic time is an indicative one, at which the virus is displaced. We also impose the condition that $\psi_\alpha(\tau)$ has the divergent first moment, which requires the exponent $\alpha$ to be in the interval $(0, 1)$. Eq. (6) implies that $\psi_\alpha(\tau)$ decays as a power law like, $\psi_\alpha(\tau) \sim s^\alpha / \tau^{1+\alpha}$, for the long time step, $\tau \gg s$, as mentioned earlier.

We expand $\tilde{f}$ up to the second order of $\Delta$ after substituting Eq. (5) into Eq. (4). Then, neglecting the term $<\Delta^2> \int d\alpha\, P(\alpha)(su)^\alpha$ with $<\Delta^2> \equiv \int_{-\infty}^{\infty} d\Delta\, \Delta^2 \phi(\Delta)$ ($u$ being small in the long time behavior), we have

$$\tilde{f}(x,u) = \frac{<\Delta^2>}{2 \int d\alpha\, P(\alpha)(su)^\alpha} \frac{\partial^2 \tilde{f}(x,u)}{\partial x^2} + \delta(x)\frac{1}{u}. \tag{7}$$

Performing the inverse Laplace transform of Eq. (7), we obtain the following generalized fractional diffusion equation:

$$\int d\alpha\, P(\alpha)\, s^{\alpha-1}\, {}_0\mathcal{D}_t^{-(1-\alpha)} \frac{\partial f(x,t)}{\partial t} = D\frac{\partial^2 f(x,t)}{\partial x^2}, \tag{8}$$

where the diffusion constant, $D$, is calculated to be $D = <\Delta^2>/(2s)$ and a mathematical fact of fractional operator [10], $\mathcal{L}({}_0\mathcal{D}_t^{-\alpha} g(x,t))(u) = u^{-\alpha}\tilde{g}(x,u)$, has been used. For the virus in a given local block, taking $P(\alpha) = \delta(\alpha - \alpha_0)$ with a certain exponent, $\alpha_0 \in (0,1)$, in Eq. (8), and applying the operator, $(\partial/\partial t)\, {}_0\mathcal{D}_t^{-\alpha_0}$, to Eq. (8), we obtain traditional fractional kinetic theory [10]. Accordingly, the mean square displacement turns out to have the form in Eq. (1), reproducing the behavior observed in the experiments. It is noted that normal diffusion is realized in the limit, $\alpha_0 \to 1$, of the present theory. We also mention that substituting this distribution of the fluctuations into Eq. (2), taking not $\psi_{\alpha_0}(\tau)$ but $\psi(\tau) = \delta(\tau - \tau_0)$ with a finite time step, $\tau_0$ (i.e., the deterministic case) as the distribution of $\tau$ in Eq. (3), Eq. (2) becomes reduced to the basic equation in Einstein's approach to Brownian motion [15] after the replacement, $t \to t + \tau_0$.

Thus, we have formulated the generalized fractional kinetic theory by introducing exponent fluctuations in view of superstatistics.

Equation (8) is formal without determining the distribution $P(\alpha)$. In the next section, a proposition for it is presented.

## 3. Fluctuations of the exponent and the maximum entropy principle

There are some discussions about determination of temperature fluctuations in superstatistics in the literature (see, for example, Refs. [20-22]). There, the maximum entropy principle plays a key role for deriving the distribution of temperature fluctuations. As we have seen above, it is necessary to clarify the statistical property of exponent fluctuations. Below, following the discussion in Ref. [9], we propose the distribution $P(\alpha)$ based on not only the experimental data but also the maximum entropy principle.

The result of the analysis of the trajectories of 104 viruses in Ref. [1] is as follows: in the form of Eq. (1), 53 trajectories among them show $\alpha = 1$, whereas other 51 exhibit $\alpha$ varying between 0.5 and 0.9. Besides this fact, there is no further information available about the weights of $\alpha \in (0.5, 0.9)$. The exponent for the free virus might differ from that for the virus contained in the endosome, in general. However, we here assume that the exponents found in both the free and endosomal forms are slightly different from each other. According to the result, normal diffusion is often to be realized. On the other hand, subdiffusion with the exponent near $\alpha = 0$ may seldom be the case, since the virus tends to reach the nucleus of the cell [1]. From these considerations, we propose the following Poisson-like form of fluctuations [9]:

$$\hat{P}(\alpha) \propto e^{\lambda \alpha} \qquad (9)$$

with a positive constant, $\lambda$.

Before proceeding, we develop a further discussion about the distribution $\hat{P}(\alpha)$ in Eq. (9). This distribution can be derived by the maximum entropy principle. To see this, let us recall that the cytoplasm is regarded as a medium being imaginarily divided into a lot of small blocks. In other words, the cytoplasm is thought of as a collection of these blocks. In this sense, introducing entropy associated with the fluctuations of the exponent for a block in the medium, we show how the distribution in Eq. (9) is derived from the maximum entropy principle.

Our situation here is the case when no information is available about how the exponent locally distributes over the cytoplasm. Accordingly, we consider all of possible distinct collections in terms of the local fluctuations. In the case of discrete values of the exponent, we form each collection by constructing the blocks with a set of different values of the exponent, $\{\alpha_i\}_i$. In terms of the statistical property of the fluctuations, no difference exists in these collections. That is, the fluctuations in a given collection are statistically equivalent to those in other collections but are locally not.

Now, as the entropy associated with the fluctuations, we evaluate the following quantity:

$$S = \frac{\ln G}{N}. \qquad (10)$$

Here, $N$ is the total number of the blocks in the medium, and $G = N! / \prod_i n_{\alpha_i}!$ with $n_{\alpha_i}$ being the number of the blocks with the exponent, $\alpha_i$, in the medium is the total number of the distinct collections. We suppose that $N$ and $n_{\alpha_i}$'s are large since the medium consists of many blocks. Using the Stirling approximation [i.e., $\ln(M!) \cong M \ln M - M$ for large $M$], we write the quantity in Eq. (10) in the form of the Shannon entropy, $S \cong -\sum_i P_{\alpha_i} \ln P_{\alpha_i}$, where $P_{\alpha_i} = n_{\alpha_i}/N$ is the probability of finding the exponent $\alpha_i$ in a given block of the medium. Correspondingly, we express the entropy in the case of continuous values of the exponent as follows:

$$S[P] = -\int d\alpha \, P(\alpha) \ln P(\alpha), \qquad (11)$$

where we have replaced the sum over $i$ by integral over $\alpha$.

Let us derive the distribution in Eq. (9) based on the maximum entropy principle with the Shannon entropy in Eq. (11). Our situation is the case when information is only available about the statistical property of the fluctuations. In this case, maximization condition of $S[P]$ in Eq. (11) with respect to $P(\alpha)$ under the constraints on both the normalization condition, $\int d\alpha\, P(\alpha) = 1$, and the expectation value of $\alpha$, $\int d\alpha\, P(\alpha)\alpha = \bar{\alpha}$, reads

$$\delta_P \left\{ S[P] - \kappa \left( \int d\alpha\, P(\alpha) - 1 \right) + \lambda \left( \int d\alpha\, P(\alpha)\alpha - \bar{\alpha} \right) \right\} = 0. \qquad (12)$$

Here, $\kappa$ and $\lambda$ are the Lagrange multipliers associated with the constraints on the normalization condition and the expectation value, respectively, and $\delta_P$ stands for the variation with respect to $P(\alpha)$. Since it is naturally supposed that the virus tends to reach the nucleus, we have imposed the condition, $P(1) > P(0)$, which requires $\lambda$ to be a positive Lagrange multiplier, in Eq. (12). The solution of Eq. (12) is given by $\check{P}(\alpha) \propto e^{\lambda\alpha}$, showing the distribution in Eq. (9).

Substitution of the distribution $\hat{P}(\alpha)$ in Eq. (9) into Eq. (8) reads the generalized fractional diffusion equation with the Poisson-like form of fluctuations [9].

Thus, we see that the present framework yields the generalized fractional kinetic theory for describing the infection pathway of the virus in the cytoplasm.

4. **Comment on scaling nature of the motion of the virus**

In this section, we present a brief discussion about the motion of the virus within the present framework. Below, it is shown that the motion of the virus may obey a scaling law, which is not studied in the previous work [9].

Using $\hat{P}(\alpha)$ in Eq. (9) in Eq. (2), we here consider the following probability:

$$f(x,t)\,dx = dx \int_0^1 d\alpha\, \hat{P}(\alpha) f_\alpha(x,t). \tag{13}$$

Then, Eq. (7) corresponds to

$$\tilde{f}(x,u) = \frac{<\Delta^2>}{2\int_0^1 d\alpha\, \hat{P}(\alpha)(su)^\alpha} \frac{\partial^2 \tilde{f}(x,u)}{\partial x^2} + \delta(x)\frac{1}{u}. \tag{14}$$

The solution of this equation is given by

$$\tilde{f}(x,u) = \frac{1}{2u}\sqrt{\frac{a(1/u)}{<\Delta^2>_\lambda}} \exp\left(-\sqrt{\frac{a(1/u)}{<\Delta^2>_\lambda}}\,|x|\right) \qquad (15)$$

with $<\Delta^2>_\lambda \equiv <\Delta^2>(e^\lambda-1)/(2\lambda)$. The function, $a(y)$, appearing in this solution is defined by

$$a(y) \equiv \frac{1 - e^\lambda s/y}{\ln[y/(e^\lambda s)]}, \qquad (16)$$

which *varies slowly at infinity* [23]: $\lim_{y\to\infty}[a(\varepsilon y)/a(y)] = 1$ for any positive constant, $\varepsilon$. Therefore, we apply a Tauberian theorem for the Laplace transforms (see Theorem 4, p.446 of Ref. [23]), which implies in the present case that $f(x,t)$ asymptotically behaves as $f(x,t) \sim \sqrt{a(t)/(4<\Delta^2>_\lambda)}\,\exp\left(-\sqrt{a(t)/<\Delta^2>_\lambda}\,|x|\right)$, for large elapsed time. For such a time, $a(t)$ behaves as $a(t) \sim 1/\ln[t/(e^\lambda s)]$. Accordingly, the asymptotic behavior of $f(x,t)$ for the large elapsed time, $t \gg e^\lambda s$, is seen to be

$$f(x,t) \sim \frac{1}{\sqrt{4<\Delta^2>_\lambda \ln[t/(e^\lambda s)]}} \exp\left(-\frac{|x|}{\sqrt{<\Delta^2>_\lambda \ln[t/(e^\lambda s)]}}\right). \qquad (17)$$

Since it satisfies the following scaling law:

$$f(x,t) \sim \frac{1}{\sqrt{\ln[t/(e^\lambda s)]}} \hat{f}\left(\frac{x}{\sqrt{\ln[t/(e^\lambda s)]}}\right), \qquad (18)$$

where $\hat{f}(x) \equiv \left(1/\sqrt{4<\Delta^2>_\lambda}\right) \exp\left(-|x|/\sqrt{<\Delta^2>_\lambda}\right)$ is a scaling function, the spatial spread of $f(x,t)$ such as its half width, $l$, is seen to scale as

$$l \sim \sqrt{\ln[t/(e^\lambda s)]}, \qquad (19)$$

which implies that the motion of the virus exhibits logarithmic behavior. This behavior seems to be supported well by the previous work [9], in which the mean square displacement of the virus, $\overline{x^2}$, is calculated by using Eq. (14) and is found to behave as, $\overline{x^2} \sim \ln[t/(e^\lambda s)]$, for the large elapsed time, showing that $\sqrt{\overline{x^2}}$ has the form in Eq. (19). Therefore, this observation seems to indicate that the scaling law in Eq. (18) may be established.

Finally, we wish to mention that it is an interesting question to examine if the scaling law in Eq. (18) can experimentally be observed in the infection pathway.

## 5. Conclusion

We have developed the theoretical framework for establishing a generalized fractional kinetics for the infection pathway of an adeno-associated virus in cytoplasm of a living *HeLa* cell in view of superstatistics. In this framework, the existence of a large separation of two time scales in the infection pathway is explicitly taken into account. Assuming slowness of variation of exponent fluctuations, we have proposed a Poisson-like form of the fluctuations based on the experimental data as well as the maximum entropy principle. In addition, we have found that the motion of the virus may obey a scaling law. We believe that the present theory gives novel physical insight into traditional theory of anomalous diffusion.